\documentclass[nonacm,sigconf]{acmart}

\usepackage{datetime}
\usepackage{subfig}
\usepackage{multirow}
\usepackage{nameref}

\usepackage{amsmath}
\usepackage[linesnumbered,ruled,noend]{algorithm2e}

\AtBeginDocument{%
  \providecommand\BibTeX{{%
    \normalfont B\kern-0.5em{\scshape i\kern-0.25em b}\kern-0.8em\TeX}}}

\setcopyright{none}
\renewcommand\footnotetextcopyrightpermission[1]{}
\begin{document}
\title{Reconciling Conflicting Data Curation Actions:
Transparency Through Argumentation}

\author{Yilin Xia}
\affiliation{%
  \institution{University of Illinois}
  \city{Urbana-Champaign, IL}
  \country{USA}}
\email{yilinx2@illinois.edu}

\author{Shawn Bowers}
\affiliation{%
  \institution{Gonzaga University}
  \city{Spokane, Washington}
  \country{USA}}
\email{bowers@gonzaga.edu}

\author{Lan Li}
\affiliation{%
  \institution{University of Illinois}
  \city{Urbana-Champaign, IL}
  \country{USA}}
\email{lanl2@illinois.edu}

\author{Bertram Lud\"{a}scher}
\affiliation{%
  \institution{University of Illinois}
  \city{Urbana-Champaign, IL}
  \country{USA}}
\email{ludaesch@illinois.edu}

\renewcommand{\shortauthors}{Yilin Xia, Shawn Bowers, Lan Li, and Bertram Lud\"{a}scher}

\definecolor{RoyalBlue}{rgb}{0.25, 0.41, 0.88}
\definecolor{DarkOrange}{rgb}{1.0, 0.55, 0.0}
\definecolor{DarkYellow}{rgb}{0.94, 0.8, 0.0}
\newcommand{\xblue} {{\ensuremath{\textcolor{RoyalBlue}{\mathsf{blue}}}}}
\newcommand{\xorange} {{\ensuremath{\textcolor{DarkOrange}{\mathsf{orange}}}}}
\newcommand{\xyellow}{{\ensuremath{\textcolor{DarkYellow}{\mathsf{yellow}}}}}

\definecolor{lightorange}{RGB}{255, 195, 128}
\definecolor{lightblue}{RGB}{155, 223, 251}
\definecolor{lightgreen}{RGB}{34, 127, 34}
\newcommand{\xlightblue} {{\ensuremath{\textcolor{lightblue}{\mathsf{light~blue}}}}}

\newcommand{\xlightorange} {{\ensuremath{\textcolor{lightorange}{\mathsf{light ~orange}}}}}

\newcommand{\xlightgreen} {{\ensuremath{\textcolor{lightgreen}{\mathsf{light~green}}}}}

\newcommand{\bertram}[1]{\textcolor{orange}{Bertram:#1}}
\newcommand{\shawn}[1]{\textcolor{purple}{Shawn: #1}}
\newcommand{\yilin}[1]{\textcolor{brown}{Yilin: #1}}
\newcommand{\lan}[1]{\textcolor{cyan}{Lan: #1}}

\begin{abstract}
  We propose a new approach for modeling and reconciling conflicting data cleaning actions. Such conflicts  arise naturally in collaborative data curation settings where multiple experts work independently and then aim to put their efforts together to improve and accelerate data cleaning. The key idea of our approach is to model conflicting updates as a formal \emph{argumentation framework} (AF). Such argumentation frameworks  can be automatically analyzed and solved by translating them to a logic program $P_{AF}$ whose declarative semantics yield a transparent solution with many desirable properties, e.g., uncontroversial updates are accepted, unjustified ones are rejected, and the remaining ambiguities are exposed and presented to users for further analysis. After motivating the problem, we introduce our approach and illustrate it with a detailed running example introducing both well-founded and stable semantics to help understand the AF solutions. We have begun to develop open source tools and Jupyter notebooks that demonstrate the practicality of our approach. In future work we plan to develop a  toolkit for conflict resolution that can be used in conjunction with OpenRefine, a popular interactive data cleaning tool.
\end{abstract}


\maketitle

\section{Introduction}

Data curation and data wrangling are critically important, labor-intensive, and error-prone phases in data science. A popular claim is that about 80\% of the effort involved in data analysis projects is spent on cleaning and preparing data sets \cite{dasu_exploratory_2003,wickham_tidy_2014}, while the subsequent analytical techniques often only constitute 20\% of the effort.
Not surprisingly, researchers and curators spend significant amounts of their time cleaning data, either with general purpose tools (e.g., Excel) and programming languages (e.g., Python, R), or using specialized tools such as OpenRefine \cite{verborgh2013using} or Wrangler \cite{kandel2011wrangler}. 

A data cleaning \emph{recipe} is a workflow $W$ that describes the data cleaning \emph{actions} (i.e., data transformations) that are performed on a ``dirty" dataset $D$ to improve its data quality and obtain a cleaner version $D'=W(D)$. Data analysis results are generally considered more trustworthy if the analysis pipeline---including the data cleaning workflow $W$---are \emph{transparent} and \emph{reproducible}. The state-of-the-art approach to increase transparency is to capture \emph{provenance} information, preferably during the whole data-lifecyle, from  data collection, through data wrangling, analysis, all the way to the scholarly publication and the creation of shared, digital research objects. In prior work, e.g., \cite{parulian2022dcm, parulian2023trust, li2019towards}, the value of prospective, retrospective, and hybrid provenance (i.e., combining the other two) has been demonstrated.

\newcommand{\mypara}[1]{\smallskip\textbf{#1.}}

\mypara{Collaborative Data Cleaning: A New Curation Challenge} In this paper, we consider the increasingly important setting where multiple researchers and curators work \emph{collaboratively} on cleaning a dataset \cite{parulian_2022}. For example, a dataset  $D$ might be split $k$-ways \emph{horizontally}, i.e.,  $D=D_1\cup\cdots\cup D_k$, based on a meaningful selection condition.\footnote{An ecology dataset, e.g., may be split by species and then assigned to different domain experts.} Another way to split the work and avoid {merge conflicts} is a \emph{vertical} split, i.e., where experts are assigned specific columns (attributes) to work on. However, there are several reasons why data curation tasks cannot always be so neatly divided up: First, there are update operations that apply to disjoint regions (rows or columns) of a dataset, yet indirectly depend on each other, e.g., via logic dependencies such as \emph{foreign keys}. We will not consider such indirect dependencies here (but explore them in future work). Another important use case involves the  assignment of overlapping regions of $D$ to multiple curators, e.g., because a clear (horizontal or vertical) cut is difficult to make, or because the overall data cleaning process can benefit from the \emph{diversity of expertise}, in which case overlapping assignments are even desirable.

\mypara{Resolving Conflicts Transparently Through Argumentation}
It is easy to see that in collaborative settings, two update actions $A$ and $B$ can be in \emph{conflict}: e.g., an existing value $v_1$ might be updated to $v_2$ by $A$ but to a different value $v_3$ by $B$. Clearly the actions $A$ and $B$ are mutually exclusive.  Asymmetric conflicts can also arise: If an update $A$ applies to a row that another action $C$ is deleting (for good reasons), then one could argue that $A$ should be rejected, since the update through $A$ is moot because the row no longer exists. 
In the following we propose to model conflicting data cleaning actions $A,B,C,\dots$, as \emph{arguments} in a formal \emph{argumentation framework} (AF) \cite{dung1995acceptability}. Such argumentation frameworks  can be automatically analyzed and solved by translating them to a logic program $P_{AF}$ whose declarative, well-founded semantics \cite{van1991well} yields a \emph{transparent} solution
with many desirable properties: e.g.,  uncontroversial updates are \emph{accepted}, unjustified ones are \emph{rejected}, and the remaining \emph{ambiguities} are exposed and presented to the user for further analysis e.g., via stable models \cite{gelfond1988stable} and conflict resolution. If a researcher, curator, or auditor questions why certain updates have been accepted, while others have been rejected, the underlying AF solution, enhanced with a game-theoretic provenance semantics \cite{ludascher2023arg} can be explored (interactively if desired) to provide a transparent, logical justification.  In the following, we briefly review some background, then 
introduce our approach and illustrate it with a detailed running example.
In the final section, we summarize and discuss plans for future work.

\section{Background \& Preliminaries}

\newcommand{\posx}[1]{\ensuremath{\mathsf{#1}}}

An \emph{argumentation framework} \posx{AF}
is a finite, directed graph $G_\posx{AF} = (V, E)$, whose vertices $V$
denote atomic \emph{arguments} and whose edges
$E \subseteq V \times V$ denote a binary \emph{attacks} relation.  An
edge $(x,y) \in E$ states that argument $x$ {\em attacks} argument
$y$. 

\begin{figure}[h!]
  \centering
  \subfloat[\emph{Attack} graph]{
    \includegraphics[width=.4\columnwidth]{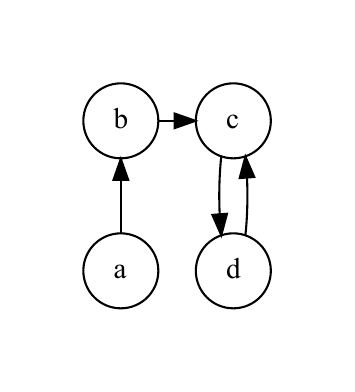}\label{fig:simple_plain}}
    \hspace{0.4cm}
  \subfloat[\emph{Grounded} extension]{
    \includegraphics[width=.4\columnwidth]{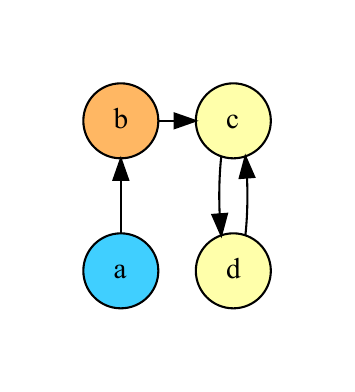}\label{fig:simple_wf}}
    \hspace{0.4cm}
    \\
  \subfloat[\emph{Stable} extension \#1]{
     \includegraphics[width=.4\columnwidth]{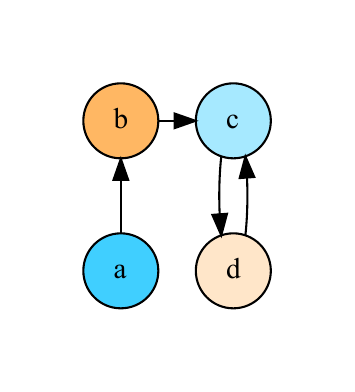}\label{fig:simple_stb1}}
  \hspace{0.4cm}
  \subfloat[\emph{Stable} extension \#2]{
    \includegraphics[width=.4\columnwidth]{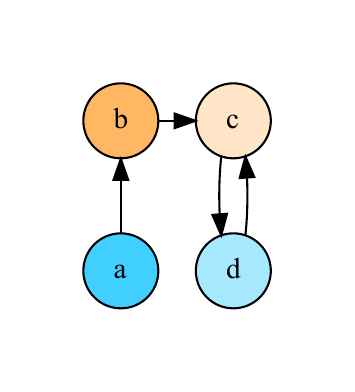}\label{fig:simple_stb2}}

  \caption{(a) AF with four arguments \posx a, \posx b, \posx c, \posx d and their \emph{attack} relation. (b) The unique, 3-valued \emph{grounded} solution: \posx a is \emph{accepted} (\xblue), \posx b is \emph{defeated} (\xorange), and \posx c, \posx d are \emph{undecided} (\xyellow). $G_\posx{AF}$ has two \emph{stable} solutions: The undecided argument \posx c can be chosen as accepted and \posx d as defeated, as in (c), or vice versa as in (d), yielding two separate stable solutions.}
  \label{fig:simple_example}
\end{figure}

\noindent An example \posx{AF} consisting of four arguments (vertices) $V=\{\posx a, \dots, \posx d\}$ and 
an \emph{attack} relation $E$ (directed edges) 
is shown in Figure~\ref{fig:simple_example}.
A subset $S \subseteq V$ of acceptable arguments is called an
\emph{extension} (or \emph{solution}), provided $S$ satisfies certain conditions.
An extension $S$ is said to \emph{attack} an argument $x$ if an
argument $y\in S$ attacks $x$. The \emph{attackers} of $S$ are the
arguments that attack at least one argument in $S$.  An extension $S$
is \emph{conflict-free} if no argument in $S$ attacks another argument
in $S$. Conversely, an extension $S$ \emph{defends} an argument $x$
if it attacks all attackers of $x$. The arguments \emph{defended by}
$S$ are those that $S$ defends; this is often described via  the
\emph{characteristic function} of an argumentation framework. 
 \cite{dung1995acceptability} and others have defined various
\emph{extension semantics}.  We consider the skeptical \emph{grounded extension} semantics, which has several advantages, e.g., it can be efficiently computed, always yields a unique, 3-valued model in which arguments (and thus edit actions) are either \emph{accepted}, \emph{rejected}, or flagged as \emph{undecided}. We will also consider \emph{stable extensions}, i.e., 2-valued solutions that refine the grounded solution by choosing acceptance or defeat of arguments in certain ways \cite{baroni_handbook_2018}.  

The overall idea and appeal of formal argumentation results from the fact that the solutions to controversial arguments can be computed automatically. As it turns out the unique well-founded model \cite{van1991well} (and the set of  stable solutions) of an argumentation framework can be obtained from a simple but powerful recursive rule:

\newcommand{\la}{\ensuremath{\leftarrow}}

\begin{equation}
  \posx{defeated}(X) \la \posx{attacks}(Y,X), \neg \, \posx{defeated}(Y). 
  \tag{$P_\posx{AF}$}
\end{equation}
The rule states that an argument $X$ is \emph{defeated} (in our terminology: a curation action is \emph{rejected}), if there exists an argument $Y$ that attacks it and that is {not} itself defeated, i.e., accepted in our data curation terminology.
Note that the \posx{AF} approach, according to \cite{dung1995acceptability}, consists of two essential components: an \emph{argument generation
  unit} (AGU) that models the arguments and the associated attack graph
(in our case the data curation actions and their conflicts), and an APU (the $P_\posx{AF}$ above) that is used to \emph{solve} an \posx{AF} and determine which arguments are accepted, rejected, and undecided, respectively. For more on formal argumentation, see the comprehensive handbook by \cite{baroni_handbook_2018}.

\begin{table}[!h]
\caption{Example dataset provided to Alice and Bob for  cleaning (``\textvisiblespace" denotes a whitespace).}
\centering
\begin{small}
\begin{tabular}{|l|l|l|}
\hline
\bf Book Title & \bf Author & \bf Date\\
\hline
\hline
Against Method & Feyerabend, P. & 1975 \\
Changing Order & Collins, H.M. & \textvisiblespace \textvisiblespace 1985 \textvisiblespace \\
Exceeding Our Grasp & P. Kyle Stanford & 2006 \\
Theory of Information &  & 1992 \\
\hline
\end{tabular}
\label{tb:raw_data}
\end{small}
\end{table}

\begin{table*}[!h]
  \caption{Data cleaning operations used by Alice and Bob (and available, e.g., in OpenRefine).}
  \label{tb:seven-operations}
  \centering
  \begin{scriptsize}
    \begin{tabular}{lp{0.5\textwidth}p{0.1\textwidth}}
      \toprule
      {\bf Data Cleaning Operation} & {\bf Description} \\
      \midrule
      
      \textsf{cell\_edit}(\textit{row\_id}, \textit{column\_name}, 
      \textit{new\_value}) & OpenRefine's single cell edit function, allowing users to hover over a cell and click ``Edit'' to modify its value. \\
      
      \textsf{del\_row}(\textit{row\_id}) & Deletes a row by using the ``Facet'' feature, selecting a relevant condition, followed by ``Remove Matching Rows''. \\
      
      \textsf{del\_col}(\textit{column\_name}) & Removes a column by going to ``Edit Column'' and selecting ``Remove this Column''. \\
      
      \textsf{split\_col}(\textit{column\_name}, \textit{separator}) & Accessed via ``Edit Column'' > ``Split into several columns'', this function splits a column into multiple ones using a specified separator and keeps the original column. \\
      
      \textsf{transform}(\textit{column\_name}, \textit{function}) & Found under ``Edit cells'' > ``Transform...'', it allows the transformation of column values using the General Refine Expression Language (GREL). \\
      
      \textsf{join\_col}(\textit{set\_of\_column\_names}, \textit{separator}, \textit{new\_column\_name}) & Combines multiple columns into a new one with a specific separator via ``Edit column'' > ``Join columns...''. \\
      
      \textsf{rename}(\textit{column\_name}, \textit{new\_column\_name}) & Rename a column under ``Edit column'' > ``Rename the column...''. \\
      \bottomrule
    \end{tabular}
  \end{scriptsize}  
\end{table*}

\begin{table*}[!h]
  \caption{Data cleaning recipes by Alice and Bob. Steps correspond to OpenRefine operations.}
  \label{tb:ab_recipes}
  \centering
  \begin{scriptsize}
    \begin{tabular}{cll}
      \toprule
      {\bf Step} & {\bf Alice's Data Cleaning Steps} & {\bf Effects of the Data Cleaning Operations} \\
      \midrule
      E & \textsf{rename}({\texttt{"}Book Title\texttt{"}, \texttt{"}Book-Title\texttt{"}}) & Replace whitespace in column name with `-' to simplify data manipulation\\
      F & \textsf{cell\_edit}(3, \texttt{"}Author\texttt{"}, \texttt{"}Stanford, P.\texttt{"}) & Edit cell value to make it consistent with the pattern from  other cells \\
      G & \textsf{transform}(\texttt{"}Date\texttt{"}, \texttt{"}value.toNumber()\texttt{"}) & Data type conversion \\
      H & \textsf{del\_row}(4) & Remove a row with a missing cell value \\
      I & \textsf{split\_col}(\texttt{"}Author\texttt{"}, \texttt{"},\texttt{"}) & Extract the lastname from the ``Author" column \\
      J & \textsf{del\_col}(\texttt{"}Author 2\texttt{"}) & Remove an unnecessary column \\
      K & \textsf{join\_col}(\texttt{"}Author 1\texttt{"}, \texttt{"}Date\texttt{"}, \texttt{"},\texttt{"} , \texttt{"}Citation\texttt{"}) & Create an in-text Citation column by combining two other columns \\
      \bottomrule
    \end{tabular}
 
\bigskip
    
    \begin{tabular}{cll}
      \toprule
      {\bf Step} & {\bf Bob's Data Cleaning Steps} & {\bf Effects of the Data Cleaning Operations} \\
      \midrule
      L & \textsf{rename}(\texttt{"}Book Title\texttt{"}, \texttt{"}Book\_Title\texttt{"}) & Replace whitespace in column name with `\_' to simplify data manipulation \\
      M & \textsf{transform}(\texttt{"}Date\texttt{"}, \texttt{"}value.trim()\texttt{"}) & Trim whitespace characters in string value \\
      N & \textsf{cell\_edit}(4, \texttt{"}Author\texttt{"}, \texttt{"}Shannon, C.E.\texttt{"}) & Add missing information \\
      O & \textsf{cell\_edit}(3, \texttt{"}Author\texttt{"}, \texttt{"}Stanford, P.K.\texttt{"}) & Edit cell value to make it consistent with the pattern from  other cells \\
      P & \textsf{split\_col}(\texttt{"}Author\texttt{"}, \texttt{"},\texttt{"}) & Extract the lastname from the ``Author" column\\
      Q & \textsf{rename}(\texttt{"}Author 1\texttt{"}, \texttt{"}Last Name\texttt{"}) & Replace the column name with a more meaningful one \\
      R & \textsf{rename}(\texttt{"}Author 2\texttt{"}, \texttt{"}First Name\texttt{"}) & Replace the column name with a more meaningful one \\
      S & \textsf{join\_col}(\texttt{"}Last Name\texttt{"}, \texttt{"}Date\texttt{"}, \texttt{"},\texttt{"} , \texttt{"}Citation\texttt{"}) & Create an in-text Citation column by combining two other columns \\
      \bottomrule
    \end{tabular}
  \end{scriptsize}  
\end{table*}

\begin{table*}[!h]
  \caption{Data cleaning results for Alice (top) and Bob (bottom).  Values depicted in \xlightgreen \ have been converted to a \posx{numeric} data type (all other columns have type \posx{string}).}
  \label{tb:ab_results}
\begin{center}
  \begin{scriptsize}
    \begin{tabular}{|l|l|l|l|l|}
        \hline

        \textbf{Book-Title} & \textbf{Author} & \textbf{Date} & \textbf{Author 1} & \textbf{Citation} \\
        \hline
        \hline
        Against Method & Feyerabend, P. & \textcolor{lightgreen}{1975} & Feyerabend & Feyerabend, 1975 \\
        \hline
        Changing Order & Collins, H.M. & \textcolor{lightgreen}{1985} & Collins & Collins, 1985 \\
        \hline
        Exceeding Our Grasp & Stanford, P. & \textcolor{lightgreen}{2006} & Stanford & Stanford, 2006 \\
        \hline
    \end{tabular}

    \bigskip
    
    \begin{tabular}{|l|l|l|l|l|l|}
        \hline
        \textbf{Book\_Title} & \textbf{Author} & \textbf{Date} & \textbf{Last Name} & \textbf{First Name} & \textbf{Citation} \\
        \hline
        \hline
        Against Method & Feyerabend, P. & 1975 & Feyerabend & P. & Feyerabend, 1975 \\
        \hline
        Changing Order & Collins, H.M. & 1985 & Collins & H.M. & Collins, 1985 \\
        \hline
        Exceeding Our Grasp & Stanford, P.K. & 2006 & Stanford & P.K. & Stanford, 2006 \\
        \hline
        Theory of Information & Shannon, C.E. & 1992 & Shannon & C.E. & Shannon, 1992 \\
        \hline
    \end{tabular}
  \end{scriptsize}  
  \end{center}
\end{table*}
\section{A Running Example for Data Cleaning}

We illustrate the key ideas of our approach to  conflict resolution in collaborative curation settings with a running example. Assume that there are two data curators, called Alice and Bob, respectively, who are working independently on  cleaning a dataset $D$. 
This dataset \cite{parulian2023trust} consists of texts in the philosophy of science, a snippet of which is shown in Table~\ref{tb:raw_data}. Each entry includes the title of the book, the author's name, and the year of publication. The task for Alice and Bob is to create a new column that adheres to the APA style guidelines for in-text citations, i.e., which require the author's last name and the year of publication.

\noindent For cleaning $D$, Alice and Bob employ several \emph{data cleaning operations} from OpenRefine. The subset of operations used by Alice and Bob is shown in Table~\ref{tb:seven-operations}, along with their parameters. These include schema-level, row-level, and cell-level operations~\cite{wfviews-idcc-2021}. Note that the {\sf split\_col} operation in OpenRefine automatically creates new columns, whereas the name of the new column created by the {\sf rename} and {\sf join\_col} operations must be explicitly given via the {\it new\_column\_name} parameter.


Alice and Bob each execute their own data cleaning \emph{recipe} (i.e., a sequence of data cleaning actions); see Table~\ref{tb:ab_recipes}. 
Unfortunately, they arrive at distinct outcomes as can be seen from the two different results in Table~\ref{tb:ab_results}. While the two result datasets are similar, there are also differences, e.g., in the Citation column. Moreover, Alice's table contains only three rows, in contrast to Bob's four as in the initial dataset (Table~\ref{tb:raw_data}). Additionally, the columns differ: Alice's version includes an ``Author 1" column, while Bob's version separates the author information into ``Last Name" and ``First Name" columns.

The key idea of our approach, described in the following sections, is to model data cleaning actions as arguments to perform the desired updates and then treat conflicting actions (like those by Alice and Bob) as arguments that can attack one another, in the sense of argumentation frameworks. By computing solutions (extensions) of the resulting argumentation frameworks, different reconciliation solutions to the conflicting recipes can be obtained automatically.

\begin{table*}[!h]
\caption{Operation Conflicts: This matrix illustrates one possible conflict (attack) relationship between pairs of data operations $A$ and $B$. For readability, the upper half of the matrix is omitted (as it can be deduced from the lower half by reversing the attack relation).}
\label{tb:atk_matrix}
\resizebox{1.8\columnwidth}{!}{%
\begin{tabular}{l|lllllll}
\hline
 &
\multicolumn{7}{c}{Operation B} \\ \hline
Operation A &
\textsf{cell\_edit}($r,c,v_2$) &
\textsf{del\_row}($r$) &
\textsf{del\_col}($c$) &
\textsf{split\_col}($c,sp_2$) &
\textsf{transform}($c,f_2$) &
\textsf{join\_col}($c,...c_j, sp_2,cn_2$) &
rename($c, c_2$) \\ \cline{1-8} 
\textsf{cell\_edit}($r,c,v_1$) &
A $\leftrightarrow$ B &
 &
 &
 &
 &
 &
 \\
\textsf{del\_row}($r$) &
A $\rightarrow$ B &
$\emptyset$ &
 &
 &
 &
 &
 \\ 
\textsf{del\_col}($c$) &
A $\rightarrow$ B &
$\emptyset$ &
$\emptyset$ &
 &
 &
 &
 \\ 
\textsf{split\_col}($c,sp_1$) &
A $\leftarrow$ B &
$\emptyset$ &
A $\leftarrow$ B &
$\emptyset$ &
 &
 &
 \\ 
\textsf{transform}($c,f_1$) &
A $\leftrightarrow$ B &
$\emptyset$ &
A $\leftarrow$ B &
A $\rightarrow$ B &
A $\leftrightarrow$ B &
 &
 \\
\textsf{join\_col}($c,...c_i,sp_1, cn_1$) &
A $\leftarrow$ B &
$\emptyset$ &
A $\leftarrow$ B &
$\emptyset$ &
A $\leftarrow$ B &
$\emptyset$ &
 \\ 
\textsf{rename}($c, c_1$) &
A $\rightarrow$  B &
$\emptyset$ &
A $\leftrightarrow$ B &
A $\rightarrow$ B &
A $\rightarrow$ B &
A $\rightarrow$ B &
A $\leftrightarrow$ B \\ \hline
\end{tabular}%
}
\end{table*}

\begin{figure*}[h]
  \centering
  \subfloat[Abstract attack relations and description of underlying data cleaning operations (cf.\ Table~\ref{tb:ab_recipes})]{
    \begin{scriptsize}
      \raisebox{0.75in}{
        \begin{tabular}{llll}
          \toprule
          \textbf{Attacks} & \textbf{Description} \\
          \midrule
          E $\leftrightarrow$ L & \textsf{rename}({\texttt{"}Book Title\texttt{"}, \texttt{"}Book-Title\texttt{"}}) $\leftrightarrow$ \textsf{rename}(\texttt{"}Book Title\texttt{"}, \texttt{"}Book\_Title\texttt{"})\\

          F ~$\leftrightarrow$ O & \textsf{cell\_edit}(3, \texttt{"}Author\texttt{"}, \texttt{"}Stanford, P.\texttt{"}) $\leftrightarrow$  \textsf{cell\_edit}(3, \texttt{"}Author\texttt{"}, \texttt{"}Stanford, P.K.\texttt{"})\\ 

          J $\leftrightarrow$ R & \textsf{del\_col}(\texttt{"}Author 2\texttt{"}) $\leftrightarrow$ \textsf{rename}(\texttt{"}Author 2\texttt{"}, \texttt{"}First Name\texttt{"})\\

          G $\leftrightarrow$ M & \textsf{transform}(\texttt{"}Date\texttt{"}, \texttt{"}value.toNumber()\texttt{"}) $\leftrightarrow$  \textsf{transform}(\texttt{"}Date\texttt{"}, \texttt{"}value.trim()\texttt{"})\\

           K $\leftarrow$ Q & \textsf{join\_col}(\texttt{"}Author 1\texttt{"}, \texttt{"}Date\texttt{"}, \texttt{"},\texttt{"} , \texttt{"}Citation\texttt{"}) $\leftarrow$ \textsf{rename}(\texttt{"}Author 1\texttt{"}, \texttt{"}Last Name\texttt{"})\\
           
           H $\rightarrow$ N & \textsf{del\_row}(4) $\rightarrow$ \textsf{cell\_edit}(4, \texttt{"}Author\texttt{"}, \texttt{"}Shannon, C.E.\texttt{"}) \\ 
          
          I $\leftarrow$ N, O & \textsf{split\_col}(\texttt{"}Author\texttt{"}, \texttt{"},\texttt{"})  $\leftarrow$ \textsf{cell\_edit}(4, \texttt{"}Author\texttt{"}, \texttt{"}Shannon, C.E.\texttt{"}), \textsf{cell\_edit}(3, \texttt{"}Author\texttt{"}, \texttt{"}Stanford, P.K.\texttt{"})\\
           
          F $\rightarrow$ P & \textsf{cell\_edit}(3, \texttt{"}Author\texttt{"}, \texttt{"}Stanford, P.\texttt{"}) $\rightarrow$ \textsf{split\_col}(\texttt{"}Author\texttt{"}, \texttt{"},\texttt{"})\\ 
          
          K $\leftarrow$ M & \textsf{join\_col}(\texttt{"}Author 1\texttt{"}, \texttt{"}Date\texttt{"}, \texttt{"},\texttt{"} , \texttt{"}Citation\texttt{"}) $\leftarrow$ \textsf{transform}(\texttt{"}Date\texttt{"}, \texttt{"}value.trim()\texttt{"})\\

          G $\rightarrow$ S & \textsf{transform}(\texttt{"}Date\texttt{"}, \texttt{"}value.toNumber()\texttt{"}) $\rightarrow$ \textsf{join\_col}(\texttt{"}Last Name\texttt{"}, \texttt{"}Date\texttt{"}, \texttt{"},\texttt{"} , \texttt{"}Citation\texttt{"})  \\
          
          \bottomrule
        \end{tabular}
      }
    \end{scriptsize}
  }
  \hfill
   \centering
  \subfloat[Argumentation Framework (solid edges) and recipe execution order (dashed edges)]{
    \includegraphics[width=1.5\columnwidth]{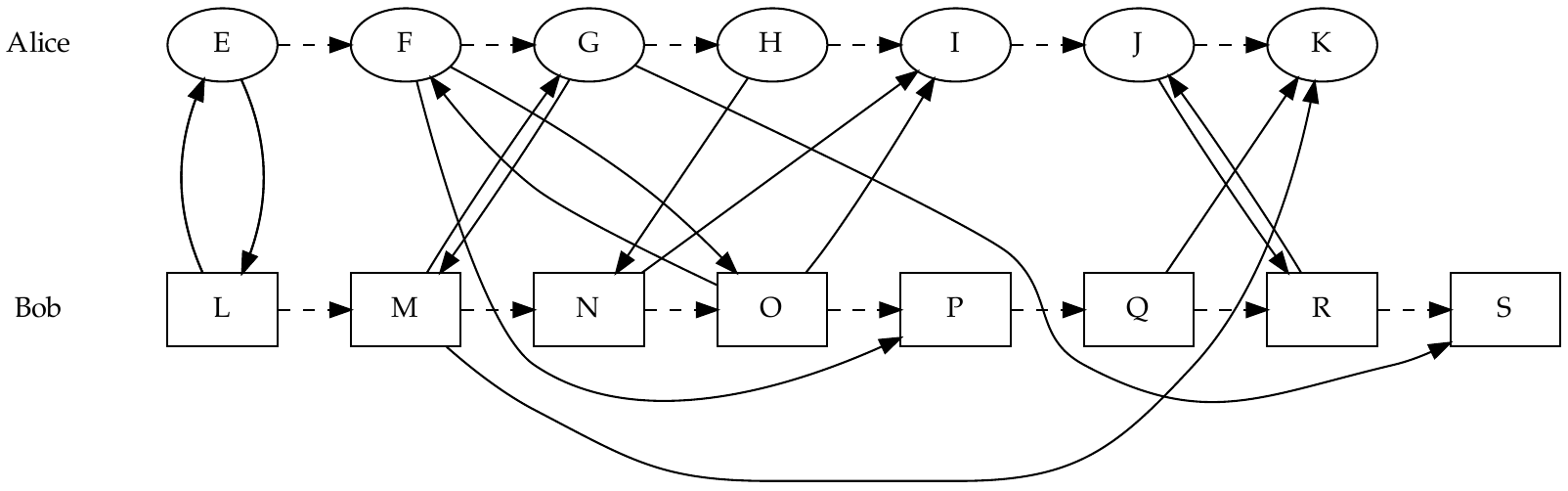}
  }
  \caption{Individual attack relations and visualized attack graph (with recipe execution order)}
  \label{fig:atk-graph}
\end{figure*}
\section{Modeling Data Cleaning Conflicts as Argumentation Frameworks}

The key idea of our approach is to treat data curation actions as \emph{arguments}, i.e., a curator claims that the corresponding operation is desirable or necessary for cleaning the data. Conflicting operations $A$ and $B$ from two different recipes are then modeled as \emph{attacks}.


 For example, $A\leftrightarrow B$ (mutual attack) means that $A$ and $B$ attack each other, so only one of them should be executed. Consider, e.g., the actions \textsf{cell\_edit}($r,c,v_1$) and \textsf{cell\_edit}($r,c,v_2$). They are considered a mutual attack whenever $v_1 \neq v_2$: Both curators agree that the cell in row $r$ and column $c$ need to be changed,  but disagree on what the new value should be. Thus \textsf{cell\_edit}($r,c,v_1$) and \textsf{cell\_edit}($r,c,v_2$) are attacking each other, denoted \textsf{cell\_edit}($r,c,v_1$) $\leftrightarrow$ \textsf{cell\_edit}($r,c,v_2$).

 Conversely, $A\rightarrow B$ means that if $A$ is accepted then $B$ is rejected (but not vice versa). For example, if a curator $C_1$ wants to delete a row $r$ and curator $C_2$ wants to edit a cell-value in $r$ (so $A=\textsf{del\_row}(r)$ and $B=\textsf{cell\_edit}(r,c,v_2)$), we could argue that $B$ should be rejected, either because it works on a cell that has been deleted already, or it performs an edit on a cell that is about to be deleted. Table~\ref{tb:atk_matrix} specifies that in such cases, deletions take priority over edits.


Another asymmetric attack relations occurs, e.g., between operations \textsf{transform}($c,f_1$) and \textsf{split\_col}($c,sp_2$). First, note that these two operations are not commutative, i.e., the result depends on the execution order. Here, for simplicity, we argue that a data cleaning transformation $f_1$ on column $c$ should take priority over a column-split operation on $c$.\footnote{Instead of rejecting the column-split operation, it might be preferable to impose an execution order, i.e., first execute $f_1$ and then split column $c$.}


By modeling the conflicts between Alice's and Bob's recipes as specified in Table~\ref{tb:atk_matrix}, we obtain the attack relation described in Figure~\ref{fig:atk-graph}(a). Additionally, this attack relationship is visualized in Figure~\ref{fig:atk-graph}(b). Note that mutual attacks are displayed using two attack edges e.g., $E\to L$ and $L\to E$. 
Operations by Alice are shown as ovals, those by Bob are depicted as boxes. Dashed lines are not attack relations but represent the \emph{execution order} of operations within a  curator's recipe.

\section{Solving AFs to Explain DC Conflicts}

After modeling data-cleaning recipes as attack graphs, the corresponding grounded and stable extensions can: ($i$) help users better understand the conflicts among the recipe actions; and ($ii$) provide guidance on how to resolve the conflicts among actions to generate one or more unified (i.e., {\em merged}) recipes. In particular, given a solved attack graph built from the recipes, we assume a merged recipe will contain the accepted actions of the corresponding attack graphs (and will not include the rejected actions). Under the grounded semantics, actions that are undecided (i.e., neither accepted nor rejected), require further analysis by users for inclusion in the merged recipe.  The stable-model semantics can then be employed to enumerate the possibilities for inclusion of the remaining (undecided) actions. Specifically, after viewing the different stable extensions, a user could select the one that they deem to be most appropriate for resolving the remaining conflicts, adding the corresponding accepted actions to create a final, merged recipe.

\begin{figure}[!ht]
  \centering
  \includegraphics[width=\columnwidth]{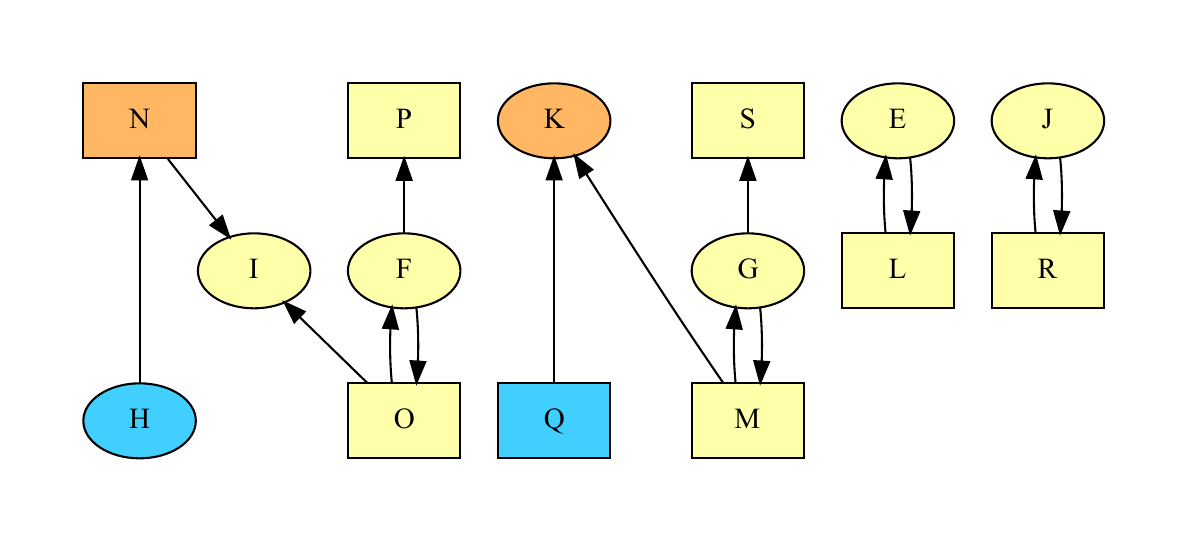}
  \caption{The grounded extension of Figure~\ref{fig:atk-graph} where (\xblue) actions are accepted, (\xorange) actions are rejected, and (\xyellow) actions are undecided.}
  \label{fig:wf_dc}
\end{figure}

As an example, Figure~\ref{fig:wf_dc} shows the solved attack graphs of Figure~\ref{fig:atk-graph} under the grounded semantics, where actions $H$ and $Q$ are accepted, $N$ and $K$ are rejected, and the remaining actions are undecided. Alice's action $H$ (deletion of row 4) is accepted because there is no other action that attacks it. Because $H$ is accepted and attacks $N$, it follows that Bob's action $N$ (to edit row 4) is rejected. Similarly, Bob's action $Q$ (to rename column ``Author 1'') is accepted because there is no other action that attacks it. Because $Q$ is accepted and attacks $K$, it follows that Alice's action $K$ (which required ``Author 1'' for a join operation) is rejected. Note, however, that this still leaves the remaining actions of Figure~\ref{fig:wf_dc} unresolved.



\begin{figure*}[h]
  \centering
  \subfloat[One of the 16 possible stable solutions]{
  \includegraphics[width=1.1\columnwidth]{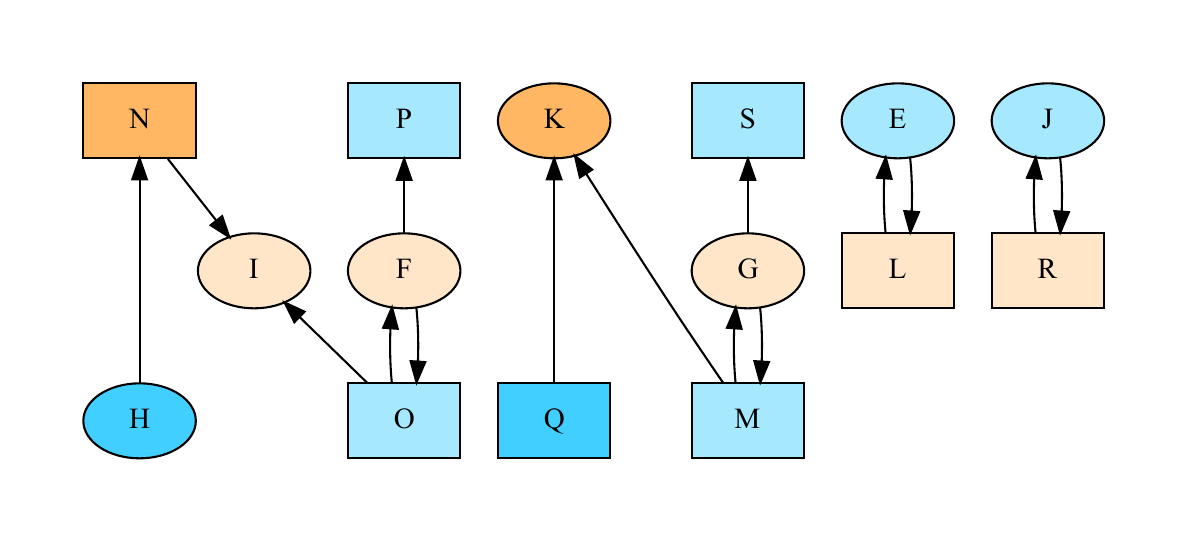}
  }
  \hfill
  \subfloat[The stable solution in sequence order according to the original recipes of Alice and Bob]{
    \includegraphics[width=1.6\columnwidth]{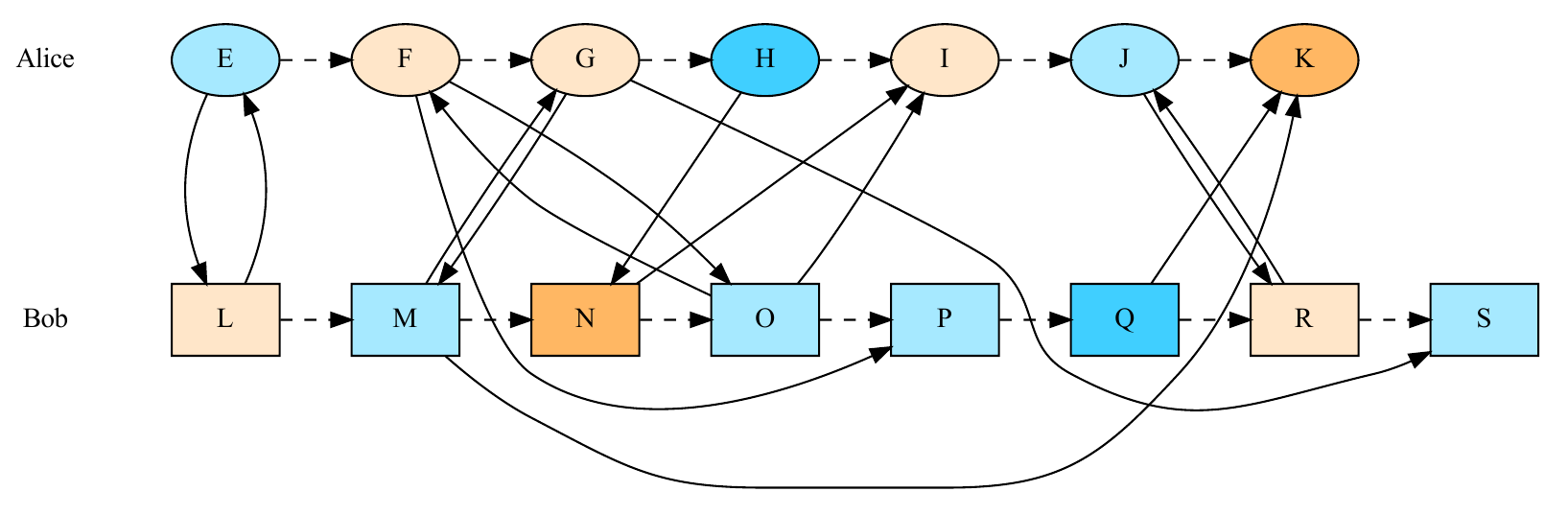}
  }
\caption{A stable extension of Figure ~\ref{fig:wf_dc} with actions appearing in \xlightblue\ being additionally accepted and those appearing \xlightorange\ additionally being rejected.}
\label{fig:stb_dc}
\end{figure*}

To help resolve the remaining conflicts, the stable models can be computed (of which there are 16 distinct solutions), one of which is shown in Figure~\ref{fig:stb_dc}(a). As shown in the figure, the stable model proposes to accept Alice's actions $E$ and $J$ along with Bob's actions $M$, $O$, $P$, and $S$.  Figure~\ref{fig:stb_dc}(b) shows the same stable model as in Figure~\ref{fig:stb_dc}(a) but displayed according to Alice's and Bob's order of actions in their respective data-cleaning recipes.

\section{Discussion \& Future Work}
\begin{table}[!t]
  \caption{One possible recipe when merging Alice and Bob's Actions.}
  \label{tb:merged_recipe}
  \centering
  \begin{scriptsize}
    \begin{tabular}{cll}
      \toprule
      {\bf Argument} & {\bf Action} & {\bf Data Curator} \\
      \midrule
      E & \textsf{rename}({\texttt{"}Book Title\texttt{"}, \texttt{"}Book-Title\texttt{"}})  & Alice \\
      M & \textsf{transform}(\texttt{"}Date\texttt{"}, \texttt{"}value.trim()\texttt{"})  & Bob \\
      H & \textsf{del\_row}(4) & Alice \\
      O & \textsf{cell\_edit}(3, \texttt{"}Author\texttt{"}, \texttt{"}Stanford, P.K.\texttt{"}) & Bob \\
      P & \textsf{split\_col}(\texttt{"}Author\texttt{"}, \texttt{"},\texttt{"}) & Bob \\
      J &  \textsf{del\_col}(\texttt{"}Author 2\texttt{"}) & Alice \\
      Q & \textsf{rename}(\texttt{"}Author 1\texttt{"}, \texttt{"}Last Name\texttt{"}) & Bob\\
      S & \textsf{join\_col}(\texttt{"}Last Name\texttt{"}, \texttt{"}Date\texttt{"}, \texttt{"},\texttt{"} , \texttt{"}Citation\texttt{"}) & Bob\\
      \bottomrule
    \end{tabular}
  \end{scriptsize}  
\end{table}

\begin{table*}[!t]
  \caption{The result of the merged recipe in Table~\ref{tb:merged_recipe} on the intial dataset in Table~\ref{tb:raw_data}.}
  \label{tb:merged_result}
  \centering
  \begin{small}
    \begin{tabular}{|l|l|l|l|l|}
        \hline
        \textbf{Book-Title} & \textbf{Author} & \textbf{Date} & \textbf{Last Name} & \textbf{Citation} \\
        \hline
        \hline
        Against Method & Feyerabend, P. & 1975 & Feyerabend & Feyerabend, 1975 \\
        \hline
        Changing Order & Collins, H.M. & 1985 & Collins & Collins, 1985 \\
        \hline
        Exceeding Our Grasp & Stanford, P.K. & 2006 & Stanford & Stanford, 2006 \\
        \hline
    \end{tabular}
  \end{small}  
\end{table*}

Assuming that the stable model of Figure~\ref{fig:stb_dc} is selected as the best resolution of the conflicting actions in Alice's and Bob's recipes, a corresponding merged recipe is shown in Table~\ref{tb:merged_recipe}. The merged recipe adheres to the order of actions in the respective recipes of Alice and Bob: $E \dashrightarrow H \dashrightarrow J$ (from Alice) and $M  \dashrightarrow O  \dashrightarrow P  \dashrightarrow Q  \dashrightarrow S$ (from Bob). Note that the specific ordering shown in Table~\ref{tb:merged_recipe} is not the only possible ordering of the actions. However, both the relative ordering of Alice's and Bob's recipe must be maintained as well as an overall ordering that generates an appropriate final data product. Specifically, in the running example, the objective is to generate a column that incorporates APA in-text citations, which in this case means that Bob's action $S$ must be the final step of the merged recipe. Moreover, there can exist additional order-based dependencies among accepted actions from different data curators. As an example, in Table~\ref{tb:merged_recipe}, Bob's action $P$ must occur before Alice's action $J$ since $P$ splits the ``Author'' column and $J$ removes ``Author 2'' (generated by the split). 



Finally, the result of applying the actions of the merged recipe in Table~\ref{tb:merged_recipe} over the initial dataset of Table~\ref{tb:raw_data} using OpenRefine is shown in Table~\ref{tb:merged_result}. 


The process of cleaning large and complex data sets can be time intensive and can require the work of multiple experts. However, conflicts can naturally arise in such collaborative data-curation settings where multiple experts work independently on the same or overlapping regions of a dataset. This paper describes an approach based on formal argumentation frameworks for modeling the actions of users' data-cleaning recipes, identifying conflicting actions across recipes, and providing users with new tools to help resolve these conflicts to generate a single, unified, merged recipe. The recipe can then be used over the original dataset to produce the final cleaned data product. By leveraging the grounded semantics of formal argumentation frameworks, it is possible to identify an initial set of accepted and rejected actions. When ambiguity is still present, the remaining actions can be resolved by selecting among one of the potentially many stable extensions. While the use of argumentation frameworks has been employed previously for resolving the justifications for specific data-cleaning actions (see \cite{santos2011using}), our work focuses on leveraging (and ultimately extending) systems such as OpenRefine, that provide a wide-range of data cleaning actions, to create new tooling for reasoning, visualizing, and automatically generating mereged data-cleaning recipes. In this paper, we have described the underlying approach through a concrete data-cleaning example, highlighting the general idea and advantages of such tools. Finally, we have begun developing open-source software and corresponding Jupyter notebooks that demonstrate the practicality of the approach \cite{xia_games-and-argumentation_2023}. In future work we plan to develop a full-featured  toolkit for conflict resolution that can be used within OpenRefine to support collaborate data cleaning projects.
\newpage
\bibliographystyle{ACM-Reference-Format}
\bibliography{sample-base}


\begin{thebibliography}{16}


\ifx \showCODEN    \undefined \def \showCODEN     #1{\unskip}     \fi
\ifx \showDOI      \undefined \def \showDOI       #1{#1}\fi
\ifx \showISBNx    \undefined \def \showISBNx     #1{\unskip}     \fi
\ifx \showISBNxiii \undefined \def \showISBNxiii  #1{\unskip}     \fi
\ifx \showISSN     \undefined \def \showISSN      #1{\unskip}     \fi
\ifx \showLCCN     \undefined \def \showLCCN      #1{\unskip}     \fi
\ifx \shownote     \undefined \def \shownote      #1{#1}          \fi
\ifx \showarticletitle \undefined \def \showarticletitle #1{#1}   \fi
\ifx \showURL      \undefined \def \showURL       {\relax}        \fi
\providecommand\bibfield[2]{#2}
\providecommand\bibinfo[2]{#2}
\providecommand\natexlab[1]{#1}
\providecommand\showeprint[2][]{arXiv:#2}

\bibitem[par(2022)]%
        {parulian_2022}
 \bibinfo{year}{2022}\natexlab{}.
\newblock \bibinfo{booktitle}{\emph{{Conceptual Model and Framework for Collaborative Data Cleaning}}}. \bibinfo{publisher}{\url{https://zenodo.org/records/6781134}}.
\newblock


\bibitem[Baroni et~al\mbox{.}(2018)]%
        {baroni_handbook_2018}
\bibfield{author}{\bibinfo{person}{Pietro Baroni}, \bibinfo{person}{Dov Gabbay}, \bibinfo{person}{Massimilino Giacomin}, {and} \bibinfo{person}{Leendert van~der Torre}.} \bibinfo{year}{2018}\natexlab{}.
\newblock \bibinfo{booktitle}{\emph{\href{https://philpapers.org/rec/BARHOF}{Handbook of {Formal} {Argumentation}}}}.
\newblock \bibinfo{publisher}{London, England: College Publications}.
\newblock


\bibitem[Dasu and Johnson({[n.\,d.]})]%
        {dasu_exploratory_2003}
\bibfield{author}{\bibinfo{person}{Tamraparni Dasu} {and} \bibinfo{person}{Theodore Johnson}.} \bibinfo{year}{[n.\,d.]}\natexlab{}.
\newblock \bibinfo{booktitle}{\emph{Exploratory Data Mining and Data Cleaning}}.
\newblock \bibinfo{publisher}{John Wiley \& Sons}.
\newblock


\bibitem[Dung(1995)]%
        {dung1995acceptability}
\bibfield{author}{\bibinfo{person}{Phan~Minh Dung}.} \bibinfo{year}{1995}\natexlab{}.
\newblock \showarticletitle{\href{https://www.sciencedirect.com/science/article/pii/000437029400041X}{On the Acceptability of Arguments and Its Fundamental Role in Nonmonotonic Reasoning, Logic Programming and n-Person Games}}.
\newblock \bibinfo{journal}{\emph{AI}} \bibinfo{volume}{77}, \bibinfo{number}{2} (\bibinfo{date}{Sept.} \bibinfo{year}{1995}), \bibinfo{pages}{321--357}.
\newblock


\bibitem[Gelfond and Lifschitz(1988)]%
        {gelfond1988stable}
\bibfield{author}{\bibinfo{person}{Michael Gelfond} {and} \bibinfo{person}{Vladimir Lifschitz}.} \bibinfo{year}{1988}\natexlab{}.
\newblock \showarticletitle{The stable model semantics for logic programming.}. In \bibinfo{booktitle}{\emph{ICLP/SLP}}, Vol.~\bibinfo{volume}{88}. Cambridge, MA, \bibinfo{pages}{1070--1080}.
\newblock


\bibitem[Kandel et~al\mbox{.}(2011)]%
        {kandel2011wrangler}
\bibfield{author}{\bibinfo{person}{Sean Kandel}, \bibinfo{person}{Andreas Paepcke}, \bibinfo{person}{Joseph Hellerstein}, {and} \bibinfo{person}{Jeffrey Heer}.} \bibinfo{year}{2011}\natexlab{}.
\newblock \showarticletitle{Wrangler: Interactive visual specification of data transformation scripts}. In \bibinfo{booktitle}{\emph{Proceedings of the sigchi conference on human factors in computing systems}}. \bibinfo{pages}{3363--3372}.
\newblock


\bibitem[Li et~al\mbox{.}(2019)]%
        {li2019towards}
\bibfield{author}{\bibinfo{person}{Lan Li}, \bibinfo{person}{Bertram Lud{\"a}scher}, {and} \bibinfo{person}{Qian Zhang}.} \bibinfo{year}{2019}\natexlab{}.
\newblock \showarticletitle{Towards more transparent, reproducible, and reusable data cleaning with OpenRefine}.
\newblock \bibinfo{journal}{\emph{iConference 2019 Proceedings}} (\bibinfo{year}{2019}).
\newblock


\bibitem[Li et~al\mbox{.}(2021)]%
        {wfviews-idcc-2021}
\bibfield{author}{\bibinfo{person}{Lan Li}, \bibinfo{person}{Nikolaus~Nova Parulian}, {and} \bibinfo{person}{Bertram Lud\"ascher}.} \bibinfo{year}{2021}\natexlab{}.
\newblock \showarticletitle{{Automatic Module Detection in Data Cleaning Workflows: Enabling Transparency and Recipe Reuse}}. In \bibinfo{booktitle}{\emph{16th International Digital Curation Conference (IDCC)}}.
\newblock
\urldef\tempurl%
\url{https://doi.org/10.5281/zenodo.5606219}
\showDOI{\tempurl}
\newblock
\shownote{\url{https://doi.org/10.2218/ijdc.v16i1.771}}.


\bibitem[Lud{\"a}scher et~al\mbox{.}(2023)]%
        {ludascher2023arg}
\bibfield{author}{\bibinfo{person}{Bertram Lud{\"a}scher}, \bibinfo{person}{Shawn Bowers}, {and} \bibinfo{person}{Yilin Xia}.} \bibinfo{year}{2023}\natexlab{}.
\newblock \showarticletitle{Games, Queries, and Argumentation Frameworks: Towards a Family Reunion}. In \bibinfo{booktitle}{\emph{7th Workshop on Advances in Argumentation in Artificial Intelligence ({AI$^3$})}}.
\newblock
\newblock
\shownote{Accepted for publication}.


\bibitem[Parulian and Lud{\"a}scher(2022)]%
        {parulian2022dcm}
\bibfield{author}{\bibinfo{person}{Nikolaus Parulian} {and} \bibinfo{person}{Bertram Lud{\"a}scher}.} \bibinfo{year}{2022}\natexlab{}.
\newblock \showarticletitle{{DCM Explorer}: a tool to support transparent data cleaning through provenance exploration}. In \bibinfo{booktitle}{\emph{14th Intl.\ Workshop on the Theory and Practice of Provenance ({TaPP})}}. \bibinfo{pages}{1--6}.
\newblock


\bibitem[Parulian and Lud{\"a}scher(2023)]%
        {parulian2023trust}
\bibfield{author}{\bibinfo{person}{Nikolaus Parulian} {and} \bibinfo{person}{Bertram Lud{\"a}scher}.} \bibinfo{year}{2023}\natexlab{}.
\newblock \showarticletitle{Trust the process: Analyzing prospective provenance for data cleaning}. In \bibinfo{booktitle}{\emph{Companion Proceedings of the ACM Web Conference 2023}}. \bibinfo{pages}{1513--1523}.
\newblock


\bibitem[Santos and Galhardas({[n.\,d.]})]%
        {santos2011using}
\bibfield{author}{\bibinfo{person}{Emanuel Santos} {and} \bibinfo{person}{Helena Galhardas}.} \bibinfo{year}{[n.\,d.]}\natexlab{}.
\newblock \showarticletitle{Using {{Argumentation}} to {{Support}} the {{User Involvement In Data Cleaning}}}. In \bibinfo{booktitle}{\emph{9th {{International Workshop}} on {{Quality}} in {{Databases}} ({{QDB}})}} ({Seattle}, 2011).
\newblock
\urldef\tempurl%
\url{http://qdb2011.dia.uniroma3.it/participants/program/index.html}
\showURL{%
\tempurl}


\bibitem[Van~Gelder et~al\mbox{.}(1991)]%
        {van1991well}
\bibfield{author}{\bibinfo{person}{Allen Van~Gelder}, \bibinfo{person}{Kenneth~A. Ross}, {and} \bibinfo{person}{John~S. Schlipf}.} \bibinfo{year}{1991}\natexlab{}.
\newblock \showarticletitle{\href{http://doi.acm.org/10.1145/116825.116838}{The Well-founded Semantics for General Logic Programs}}.
\newblock \bibinfo{journal}{\emph{J. ACM}} \bibinfo{volume}{38}, \bibinfo{number}{3} (\bibinfo{year}{1991}), \bibinfo{pages}{619--649}.
\newblock


\bibitem[Verborgh and De~Wilde(2013)]%
        {verborgh2013using}
\bibfield{author}{\bibinfo{person}{Ruben Verborgh} {and} \bibinfo{person}{Max De~Wilde}.} \bibinfo{year}{2013}\natexlab{}.
\newblock \bibinfo{booktitle}{\emph{Using {OpenRefine}}}.
\newblock \bibinfo{publisher}{Packt Publishing Ltd}.
\newblock


\bibitem[Wickham({[n.\,d.]})]%
        {wickham_tidy_2014}
\bibfield{author}{\bibinfo{person}{Hadley Wickham}.} \bibinfo{year}{[n.\,d.]}\natexlab{}.
\newblock \showarticletitle{Tidy Data}.
\newblock   \bibinfo{volume}{059} (\bibinfo{year}{[n.\,d.]}).
\newblock
Issue i10.
\urldef\tempurl%
\url{https://doi.org/10.18637/jss.v059.i10}
\showURL{%
\tempurl}


\bibitem[Xia and Lud\"ascher(2023)]%
        {xia_games-and-argumentation_2023}
\bibfield{author}{\bibinfo{person}{Yilin Xia} {and} \bibinfo{person}{Bertram Lud\"ascher}.} \bibinfo{year}{2023}\natexlab{}.
\newblock \bibinfo{title}{Games and Argumentation Demo Repository}.
\newblock
\newblock
\newblock
\shownote{\href{https://github.com/idaks/Games-and-Argumentation/tree/idcc}{github.com/idaks/Games-and-Argumentation/tree/idcc}}.


\end{thebibliography}

\end{document}